\documentclass[preprint2]{aastex}
\usepackage{amsmath}
\usepackage{amssymb}

\begin{document}

\title{Large Scale Soft X-ray Loops And Their Magnetic Chirality In Both Hemispheres}

\author{Hongqi Zhang\altaffilmark{1},
Shangbin Yang\altaffilmark{1}, Yu Gao\altaffilmark{1}, Jiangtao
Su\altaffilmark{1}, D.D. Sokoloff\altaffilmark{1,2}, K.
Kuzanyan\altaffilmark{1,3}}

\altaffiltext{1}{Key Laboratory of Solar Activity,
National Astronomical Observatories, Chinese
Academy of Sciences, Beijing 100012, China, E-mail:
hzhang@bao.ac.cn} \altaffiltext{2}{Department of Physics, Moscow
State University, Moscow 119992, Russia} \altaffiltext{3}{IZMIRAN,
Troitsk, Moscow Region 142190, Russia}

\begin{abstract}

The magnetic chirality in solar atmosphere has been studied based on
the soft X-ray and magnetic field observations.  It is found that some
 of large-scale twisted soft X-ray loop systems occur
for several months in the solar atmosphere, before the disappearance
of the corresponding background large-scale magnetic field. It
provides the observational evidence of the helicity of the
large-scale magnetic field in the solar atmosphere and the reverse
one relative to the helicity rule in both hemispheres with solar
cycles. The transfer of the magnetic helicity from the subatmosphere
is consistent with the formation of large-scale twisted soft X-ray
loops in the both solar hemispheres.
\end{abstract}


\section{Introduction}

The magnetic helicity plays an important role in the solar flares
and coronal mass ejections (CMEs), and in the dynamo processes that
causes the 11 year solar cycle (Zhang and Low, 2005). The magnetic
helicity is an integral quantity reflecting the global complexity of
the magnetic field. It can be inferred in several ways based
on the measurements of magnetic field in the solar atmosphere. The
pioneering study of magnetic helicity was taken by some authors based
on accumulation of magnetic helicity $H_m=\int{\bf A}\cdot {\bf
B}d^3x$ in the solar atmosphere (e.g. Berger and Field, 1994; Chae,
2001) and on the mean factor of force free field
$\overline{\alpha_{\rm ff}}=\overline{(\bigtriangledown\times{\bf
B})_{\bot}\cdot {\bf B}_{\bot}/{B^2_{\bot}}}$ (or the mean current
helicity $\overline{h_c}=\overline{(\bigtriangledown\times{\bf
B})_{\bot}\cdot {\bf B}_{\bot}}$) in solar active regions (e.g.
Seehafer, 1990).

It is found that mean current helicity in most of solar active
regions in the northern (southern) hemisphere tend to show the
negative (positive) sign. This trans-equatorial sign rule for the
magnetic chirality was firstly discovered by Hale et al. (1919) from
H$\alpha$ pattern of active regions. A series of studies on the
hemispheric sign rule for magnetic helicity has been presented in
recent years (Seehafer, 1990; Pevtsov et al., 1995; Abramenko et
al., 1996; Bao and Zhang, 1998; Hagino and Sakurai, 2005). It has been
noticed the reversed sign of current helicity in solar active
regions with respect to the trans-equatorial sign rule, which is found
in high correlation with the powerful solar flare-CMEs (Bao,
et al., 1999; Zhang, et al., 2000; Liu and Zhang, 2002; Wang et al.,
2004), who found that the active regions with the reversed signs of
current helicity show the higher possibility to release the magnetic
energy into the interplanetary space than the regular ones.

Statistical analysis of the observed current helicity in active
regions with solar cycle has been presented by Bao and Zhang (1998),
Zhang and Bao (1999), Bao et al. (2000), Hagino and Sakurai (2005)
and Xu et al. (2007). It has been found that the mean helicity of solar active
regions change with the phase of the solar cycle and the mean sign of
helicity occurs reversed with respect to the hemispheric rule in some phases of
the solar cycle. Several mechanisms of the magnetic helicity generation
inside of the Sun have been proposed (Longcope et al., 1998; Berger
and Ruzmaikin, 2000; Kleeorin et al., 2003; Blackman and
Brandenburg, 2003; Sokoloff et al., 2006; Zhang et al., 2006). In
comparison with the observational results, the mirror symmetrical
reverse of the magnetic helicity of solar active regions relative to
the preferred hemispheric trends at the different phases of solar
cycle have been theoretically demonstrated (Choudhuri et al, 2004;
Xu et al., 2009). Similar simulation of the distribution of
current helicity in the full-Sun has been recently developed by Yeates et al.
(2009).

Recently, the distribution of current helicity of solar active
regions with the phase of solar cycle and latitude has been
presented by Zhang et al. (2010) based on the vector magnetograms of
active regions for more than 20 years observed at Huairou Solar
Observing Station. It shows the following observational evidence:
Electric current helicity and twist follow the propagation of the
magnetic activity dynamo waves recorded by sunspots. The helicity
and twist oscillate with 11-year period like sunspots rather than
22-year period as magnetic fields do. The helicity and twist
patterns are in general anti-symmetric with respect to the solar
equator. The helicity pattern is more complicated than Hale's
polarity law for sunspots. It is found areas of the ``wrong'' sign
at the ends of the butterfly wings as well at their very beginnings.
The average amplitude of the helicity does not show any significant
dependence on the phase of solar cycle.
The maximum value of helicity, at the
surface at least, seems to occur near the edges of the butterfly
diagram of sunspots.

Using photospheric vector magnetograms from the Haleakala Stokes
Polarimeter and coronal X-ray images from the Yohkoh Soft X-Ray
Telescope (SXT), Pevtsov et al. (1997) inferred values of the
force-free field parameter $\alpha$ at both photospheric and coronal
levels within 140 active regions. They found that both values  are
well correlated. Only for active regions in which both signs of
alpha are well represented, and in which their method of analysis
therefore breaks down, both values are poorly correlated. This
implies that the helical coronal configuration of soft X-ray loops
provides the basic information on the magnetic helicity in
the solar atmosphere. It is also a considerable parameter to analyze
the distribution of hemispheric sign of magnetic helicity and its
evolution with the solar cycle.

In this paper, we present the distribution of the sign of magnetic
helicity inferred from the distribution of soft X-ray loops and its
evolution with solar cycles. We also discuss the large-scale
reversal magnetic helicity in the both hemispheres in some phase of
the solar cycle and analyze its relevance in the framework of the
solar dynamo.

\begin{figure}
\begin{center}
\includegraphics*[width=70mm,angle=0]{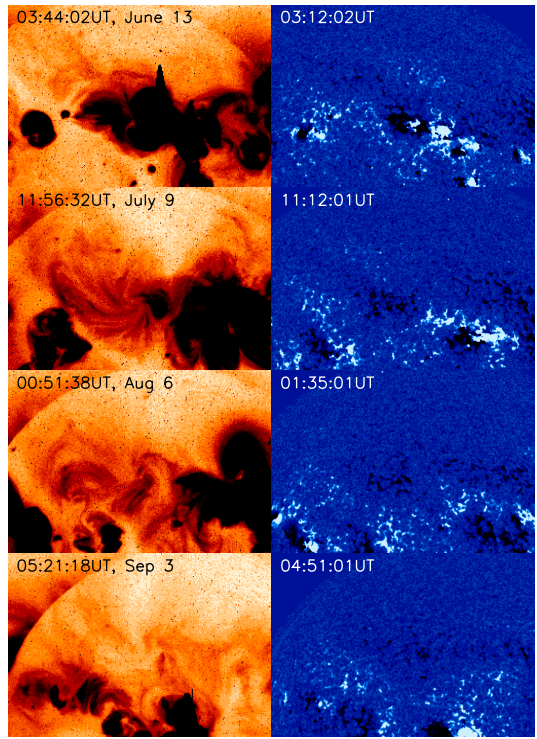}
\end{center}
\caption{Soft X-ray images of a part of the Sun (left), where the
large-scale soft X-ray loops twist clockwise, and the corresponding
photospheric magnetograms (right) in the period of 2000
June-September. The white (black) indicates the positive (negative)
polarity in the magnetograms. The top is north and right is at
west.}
\end{figure}

\begin{figure}
\begin{center}
\epsscale{0.45}
\plotone{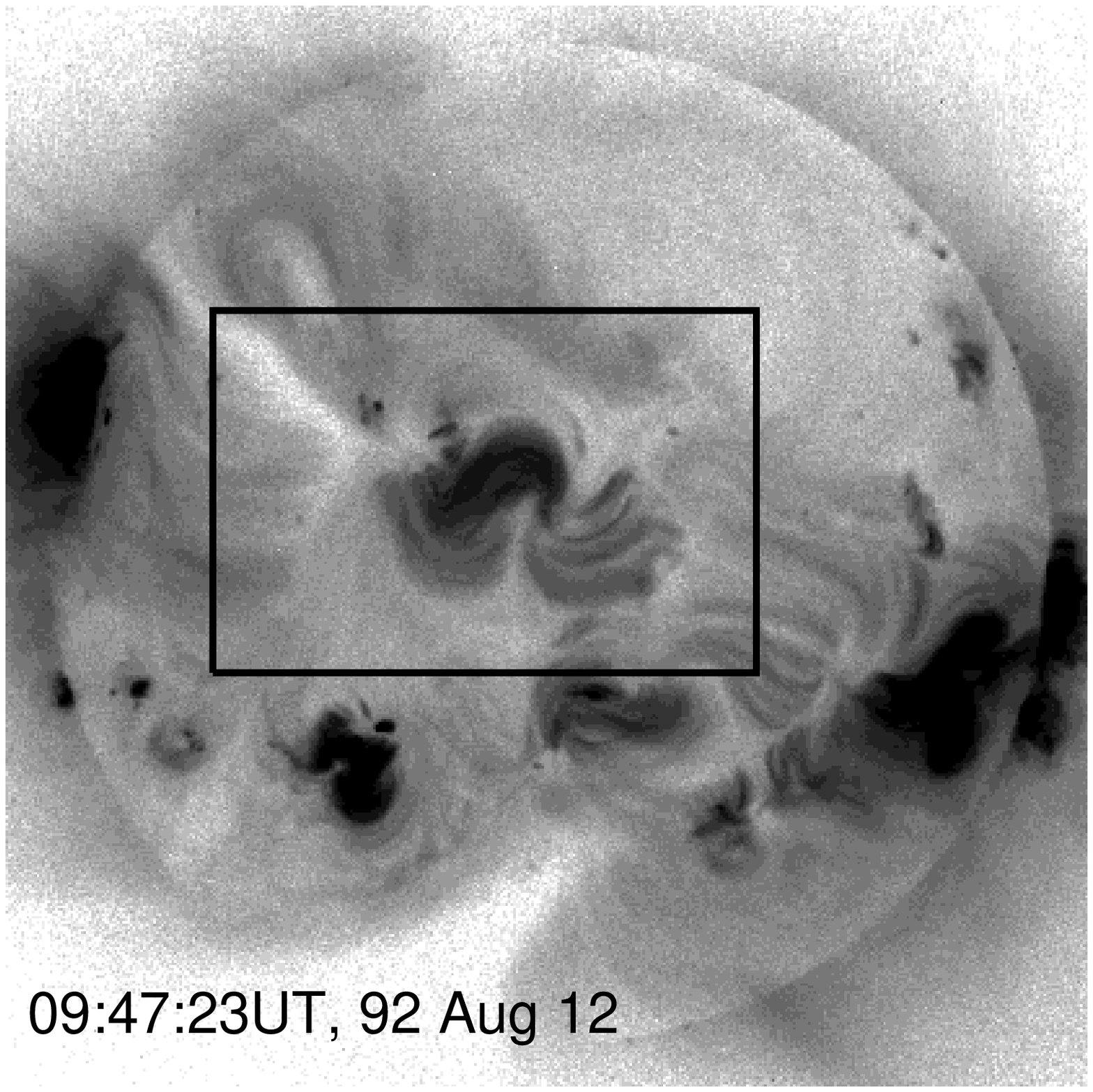}\plotone{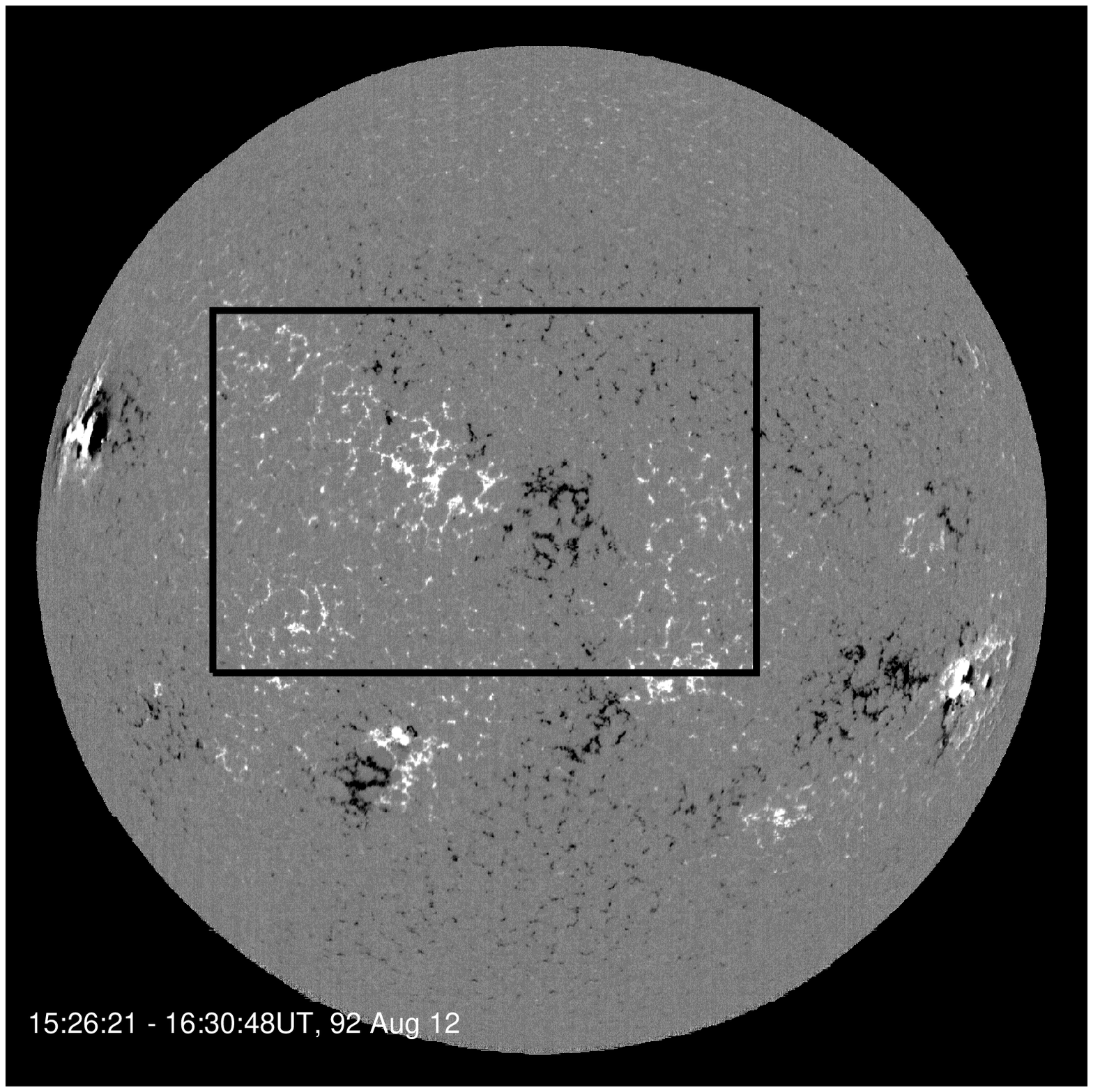} \vspace*{1mm}
\end{center}
\caption{Full disk soft X-ray image of the Sun (left) and the
corresponding photospheric magnetogram (right) on 1992 August 12.
The white (black) indicates the positive (negative) polarity in the
magnetogram. The top is north and right is at west.}
\end{figure}

\section{Magnetic chirality of soft X-ray loops related to solar
active regions}

Figure 1 shows a sample of the twisted large-scale soft X-ray
configuration in the solar northern hemisphere from the Yohkoh Soft
X-Ray Telescope (SXT). It is found that the twisted large-scale soft
X-ray configuration remained in the solar atmosphere for several
months in the period of 2000 June - September, even if the topology
of the soft X-ray configuration changed gradually. It is normally
believed that the soft X-ray configuration in the solar atmosphere
provides the basic information of magnetic field, as one believes
that the field is bound up in the ionized plasma. This large-scale
twisted magnetic field in the solar atmosphere was originated from a
decaying solar active region NOAA 9033 in 2000 June. This means that
the twisted magnetic field in the solar atmosphere is originated
from solar active regions and the diffused remainder of helical
magnetic field from the active regions can be kept a long time in
the solar atmosphere (Zhang, 2006). The twist of soft X-ray loops in
the left handedness in the solar northern hemisphere is consistent
with the handed rule of magnetic helicity presented by some authors
in recent years (Seehafer, 1990; Pevtsov et al., 1995; Abramenko et
al., 1996; Bao and Zhang, 1998; Hagino and Sakurai, 2005).

Figure 2 shows the full disk soft X-ray images of the Sun and the
corresponding photospheric magnetograms on 1992 August 12. It is
found that the right (left) handed chirality of large-scale soft
X-ray loops occurred near the center of solar disk in the northern
(southern) hemisphere. This means that the handedness of these soft
X-ray loops is opposite to the statistical handedness rule for magnetic
helicity. Figure 3 shows a series of soft X-ray images of a part of
the Sun (left), where the large-scale soft X-ray loops twist
anticlockwise, and the corresponding photospheric magnetograms
(right) in the period of 1992 June-October. This region is marked by
boxes in Figure 2. The anti-clockwise chirality does not change with evolution
of the soft X-ray configuration. This means
that the sign of helicity hold for a relative long period in the solar
atmosphere. This is also consistent with the results reported by
Zhang and Bao (1999), that the reverse helicity of solar active
regions tends to occur in some specific longitudes and holds its sign
for several solar rotations.

\begin{figure}
\begin{center}
\includegraphics*[width=70mm,angle=0]{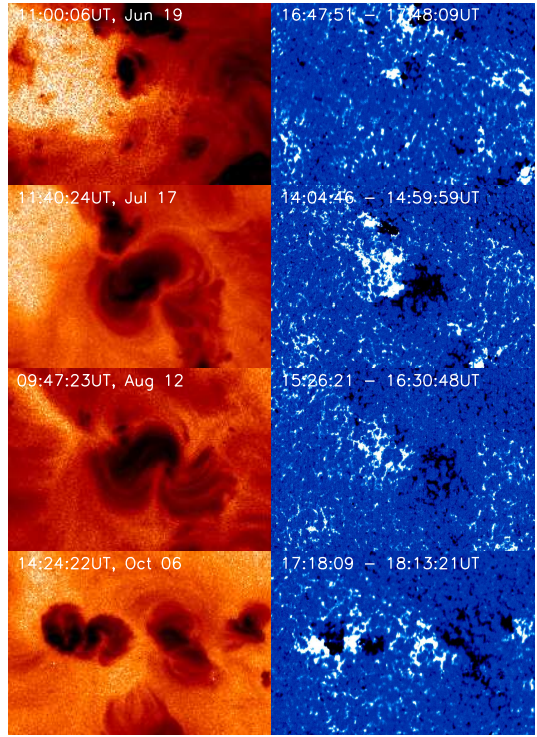}
\end{center}
\caption{Soft X-ray images of a part of the Sun (left), where the
large-scale soft X-ray loops twist anticlockwise, and the
corresponding photospheric magnetograms (right) in the period of
1992 June-October. The white (black) colors indicate the positive
(negative) polarity in the magnetograms. }
\end{figure}

Figure 4 and 5 show the other two examples of some notable
handedness of soft X-ray loops in the both hemispheres, marked by
boxes. Figure 4 shows that right handedness (positive helicity) of
the large-scale soft X-ray loops occurs in the northern hemisphere,
while Figure 5 shows left handedness (negative helicity) of the
large-scale soft X-ray loops in the southern hemisphere. These
large-scale soft X-ray loops are connected with the enhanced network
magnetic fields and active regions.

\begin{figure}
\begin{center}
\epsscale{0.45}
\plotone{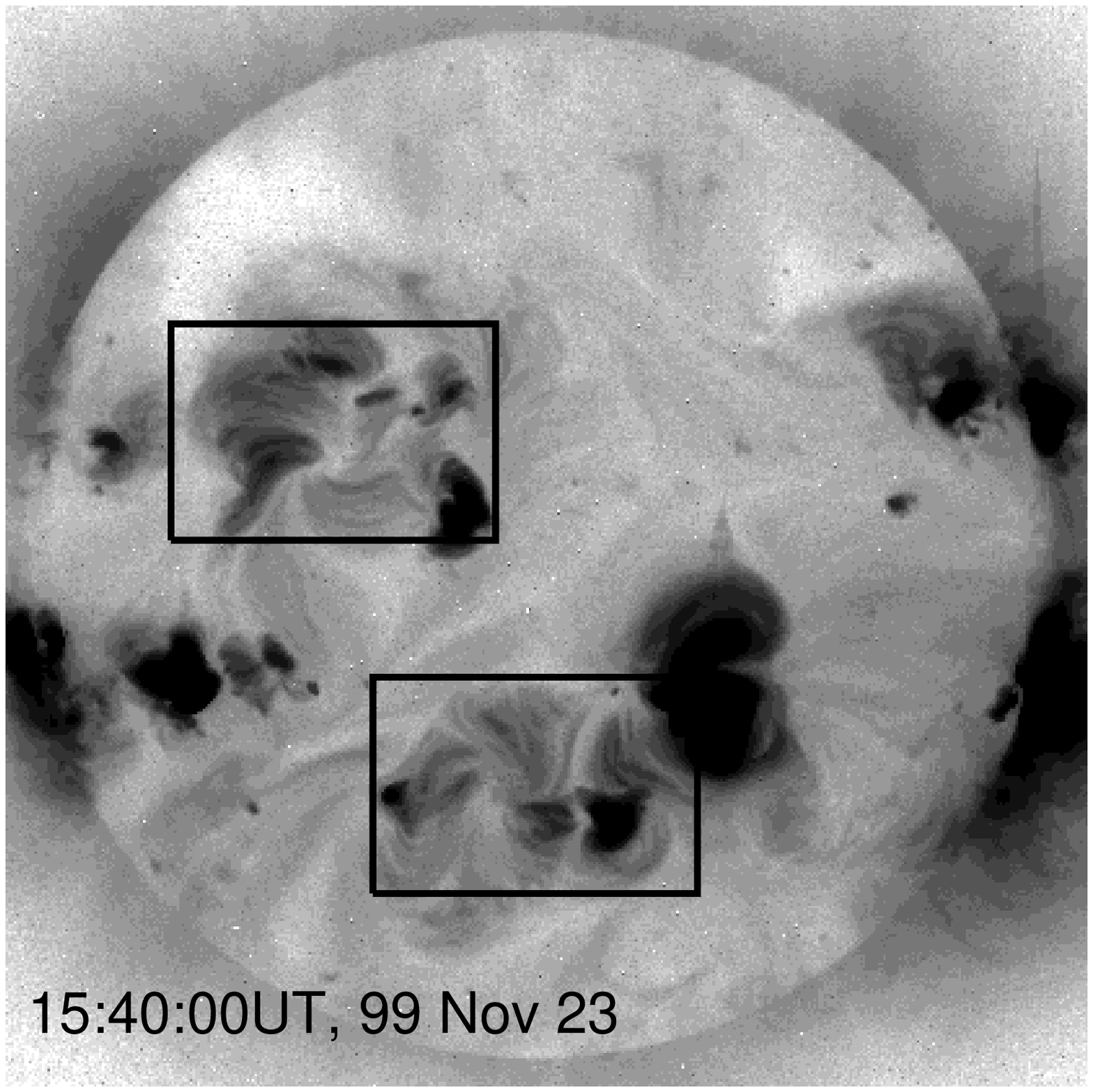}\plotone{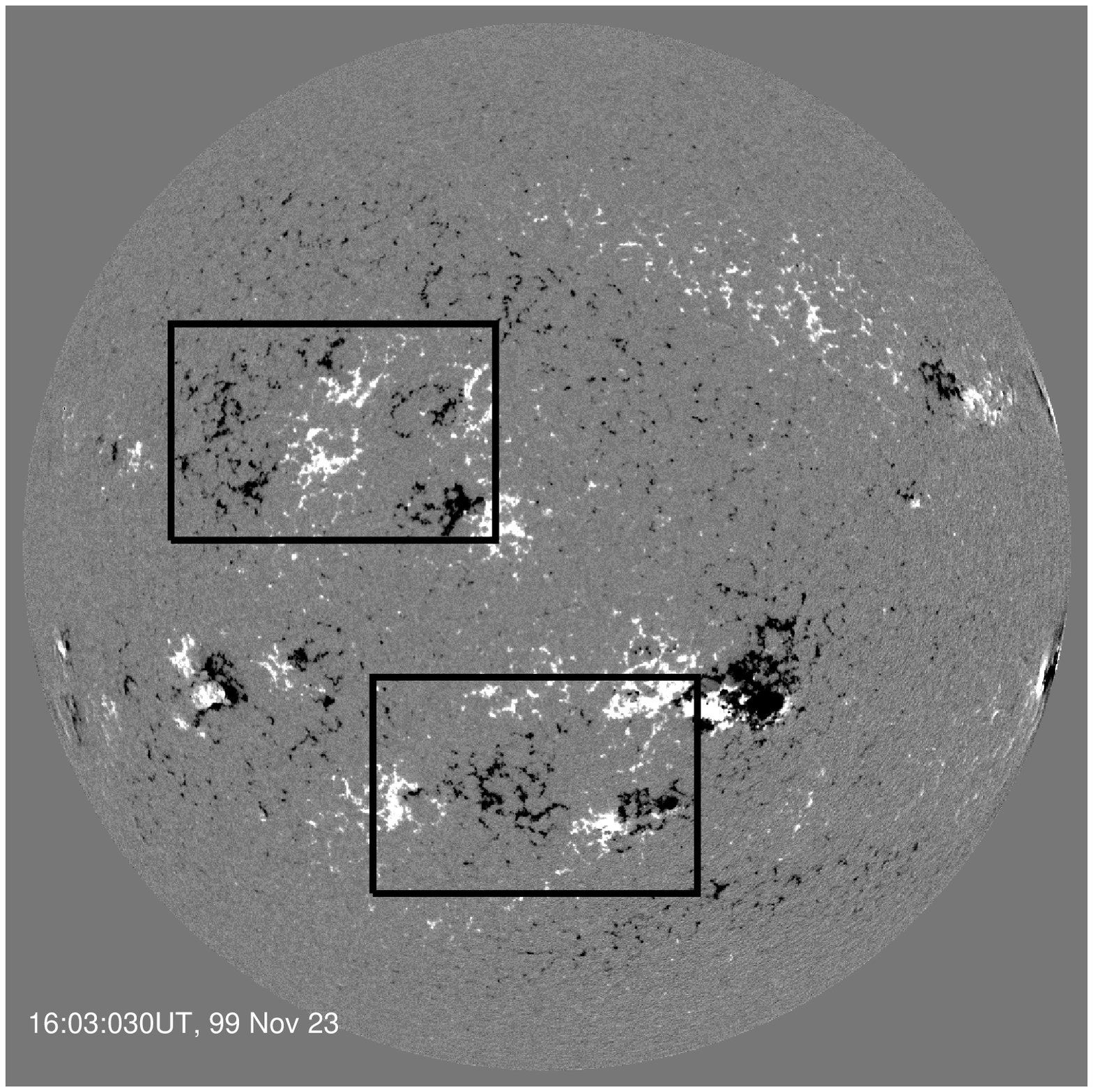} \vspace*{1mm}
\end{center}
\caption{Full disk soft X-ray image of the Sun (left) and the
corresponding photospheric magnetogram (right) on 1999 Nov 23.
The white (black) indicates the positive (negative) polarity in the
magnetogram. The top is north and right is at west.}
\end{figure}

\begin{figure}
\begin{center}
\epsscale{0.45}
\plotone{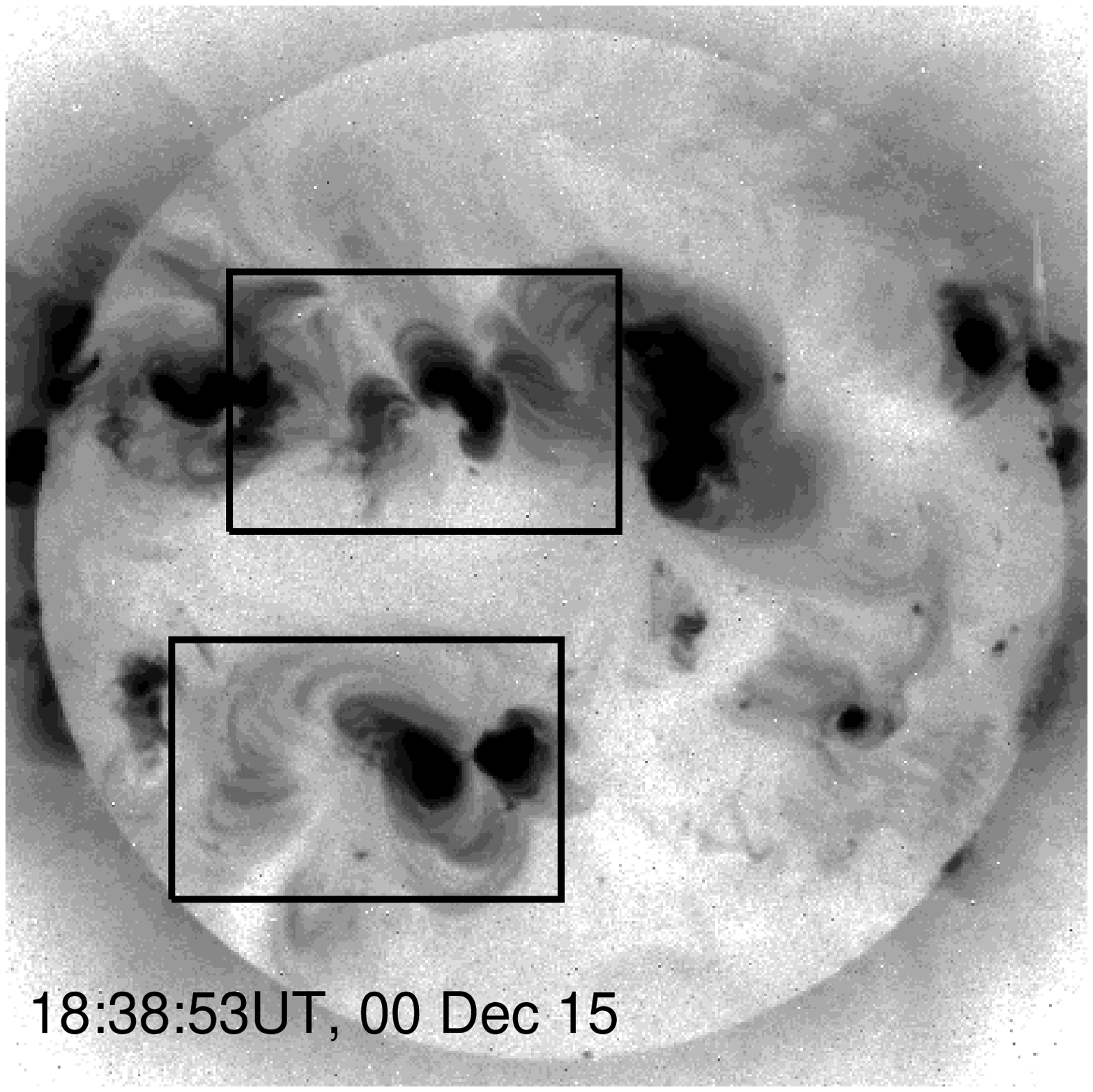}\plotone{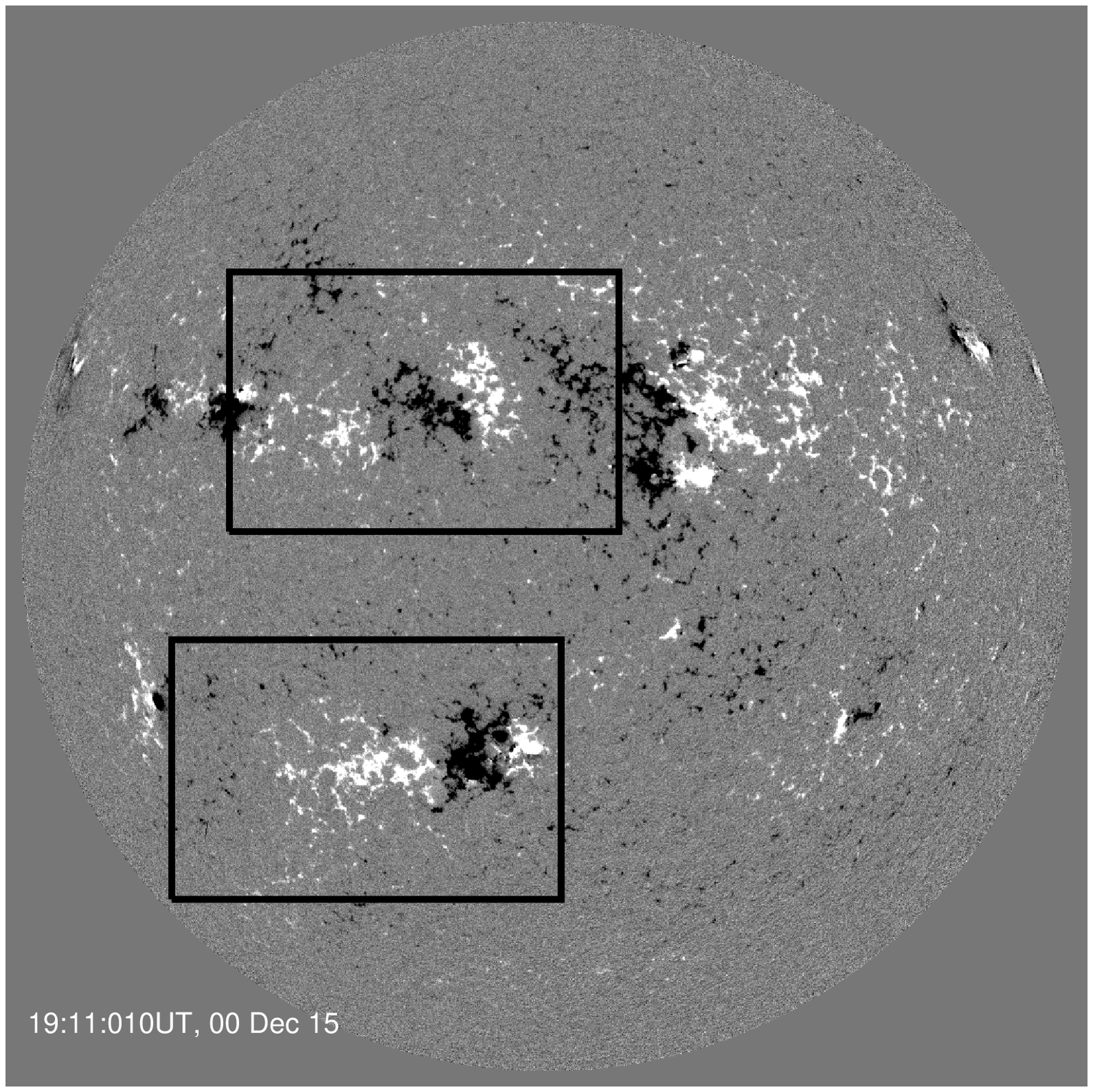} \vspace*{1mm}
\end{center}
\caption{Full disk soft X-ray image of the Sun (left) and the
corresponding photospheric magnetogram (right) on 2000 Dec 15.
The white (black) indicates the positive (negative) polarity in the
magnetogram. The top is north and right is at west.}
\end{figure}

\section{Hemispheric distribution of helical soft X-ray loops}

For analyzing the distribution of magnetic chirality in the solar
atmosphere, in Table 1 we presented statistics on 753 large-scale
soft X-ray loop systems in the period of 1991 - 2001 observed by
Yohkoh satellite. The handedness of soft X-ray loops can be inferred
by their twist or sigmoid configuration. It is found that the
handedness of soft X-ray loops statistically obeys the hemispheric
sign rule. The most of them possess left (right) handedness in the
northern (southern) hemisphere. It is found that the handedness for
about 31$\%$ soft X-ray loops can not been identified, because their
configurations are not too far from the approximation of the
potential field or can not be clearly identified as sigmoid or twist
configurations. This lack of identification does significantly not
influence the trend in the ratio of the handednesses of soft X-ray
loops between the northern and southern hemispheres. As these
unindentified soft X-ray loop systems are ignored, one can find that
the portion of the systems which are in accord with the hemispheric
rule is $77.3\%$ in the northern hemisphere and $81.5\%$ in the
southern hemisphere. It is roughly consistent with results
calculated from the vector magnetograms (Pevtsov et al., 1995; Bao
and Zhang, 1998; Hagino and Sakurai, 2005; Zhang et al., 2010).

\begin{table*}[htb]
\begin{center}
 \caption{The statistics of handedness of soft X-ray loops in the
northern and southern hemispheres\label{tbl-1}}
\begin{tabular}{ccccccccccccc}
\tableline\tableline  Year  &1991&1992&1993&1994&1995&1996&1997&1998&1999&2000&2001&Total\\
\tableline
$N n$ & 4 & 38 & 25 & 31 & 5 &5 & 16&23 &32 &24 &22 &225\\
$P n$ & 6 & 24 & 5  & 4 &3 & 1& 2& 3& 9&6 &5 &68 \\
$Q n$ & 7 & 27 & 22 & 14& 5& 5& 11& 15&18 &13 & 7 &144\\
\tableline
$P s$ & 13 & 23 & 24 & 13 & 7 & 6 & 19 & 26 & 16 & 11 & 27&185 \\
$N s$ & 7 & 8 & 10 & 4 & 2 & 1 & 2 & 2 & 2 & 3 & 1 &42 \\
$Q s$ & 3 & 12 & 4 & 5 & 7 & 4 & 5 & 22 & 14 & 6 & 7&89 \\
\tableline
 $Total$&40 & 132 & 90 &  71& 29&22 & 55 & 91 & 91 & 63 & 69 &753\\
\tableline
\end{tabular}
\tablenotetext{}{$N$ is the number of soft X-ray loops with left handedness,\\
$P$ is the number of soft X-ray loops with right handedness,\\
$Q$ is the number of unidentified soft X-ray loops.\\
The subscript $n$ and $s$ indicate the northern and southern
hemispheres, respectively.}
\end{center}
\end{table*}

Figure 6 shows the proportion of soft X-ray loops following the
hemispheric handedness rule for helicity in the northern and southern
hemispheres. It is found the change of the proportion of soft X-ray
loops following the hemispheric handedness rule of helicity and also
their imbalance of chirality in both hemispheres. The relative high tendency
of the reverse magnetic helicity has occured in 1991, 1992 and 1995 in the
northern hemisphere, while it has not been significant in the southern
hemisphere. It is consistent with the results shown in Figures 2 and 3.

Figure 7 shows the statistical latitudinal distribution of soft
X-ray loops in Figure 6. It is found the trends of the mean latitude
of soft X-ray loops to migrate towards the equator with the phase of
the solar cycles following sunspots in the butterfly diagram.
Because here were very few soft X-ray loops in 1991, 1995 and 1996
which are included in our statistics, the deviation from the
butterfly diagram in these years can be noted. The most of large
scale soft X-ray loops show the left (right) handedness in the
northern (southern) hemisphere, which follows the handedness rule
for current helicity of solar active regions, while the statistical
distribution of the reverse soft X-ray loops shows left (right)
handedness in the southern (northern) hemisphere) as one can see in
Figure 7.

\begin{figure}
\begin{center}
\epsscale{0.90} \plotone{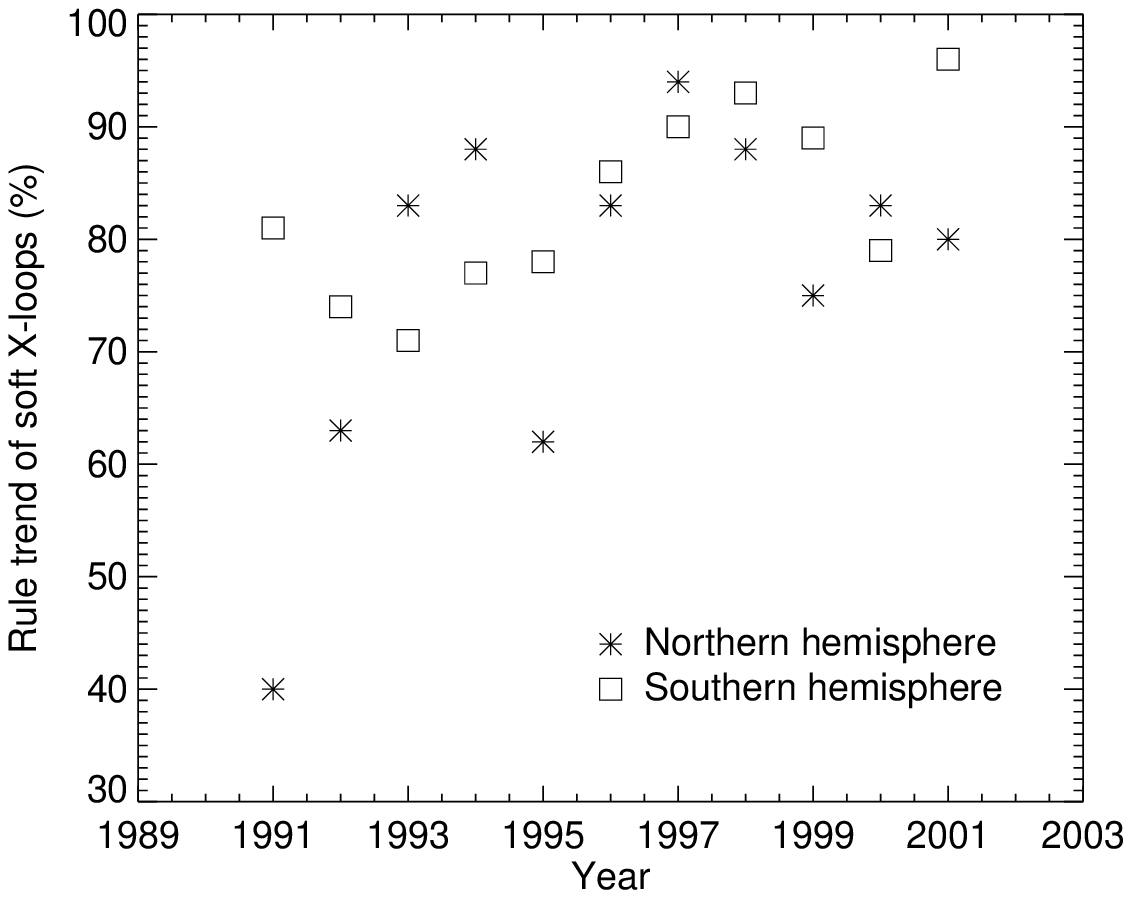}
\end{center}
\caption{The proportion of soft X-ray loops following the
hemispheric handed rule of helicity in the northern and southern
hemispheres.}
\end{figure}

\begin{figure}
\begin{center}
\epsscale{0.90} \plotone{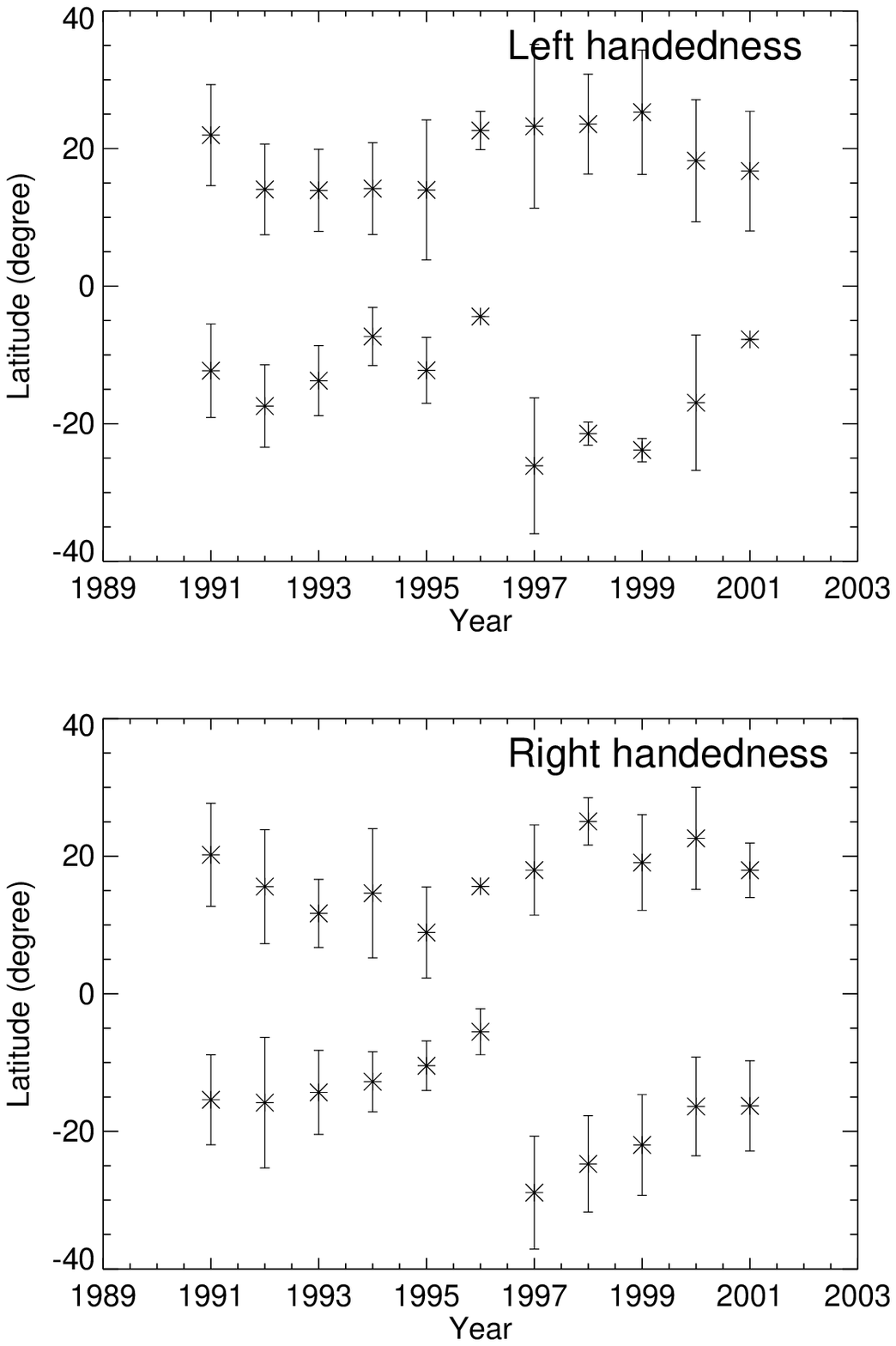}
\end{center}
\caption{The mean latitudinal distribution of soft X-ray loops with
the left and right handedness. $\sigma$-error bars are shown by vertical lines.}
\end{figure}

\section{Handedness of large-scale soft X-ray loops and magnetic (current) helicity}

It is noticed that the synthetical analysis on the accumulation of
magnetic helicity and the relationship with mean current helicity
density (and also mean force free $\alpha$) is an important way to
understand the basic information on dynamics of magnetic helicity in
solar active regions (Zhang, 2006). It can be used to analyze the
relationship between the handedness of large-scale soft X-ray loops
and the corresponding magnetic helicity in the solar atmosphere,
while the helicity of large-scale soft X-ray loops is probably
contributed from the nearby solar active regions and also enhanced
networks.

\begin{figure}
\begin{center}
\epsscale{0.90}
\plotone{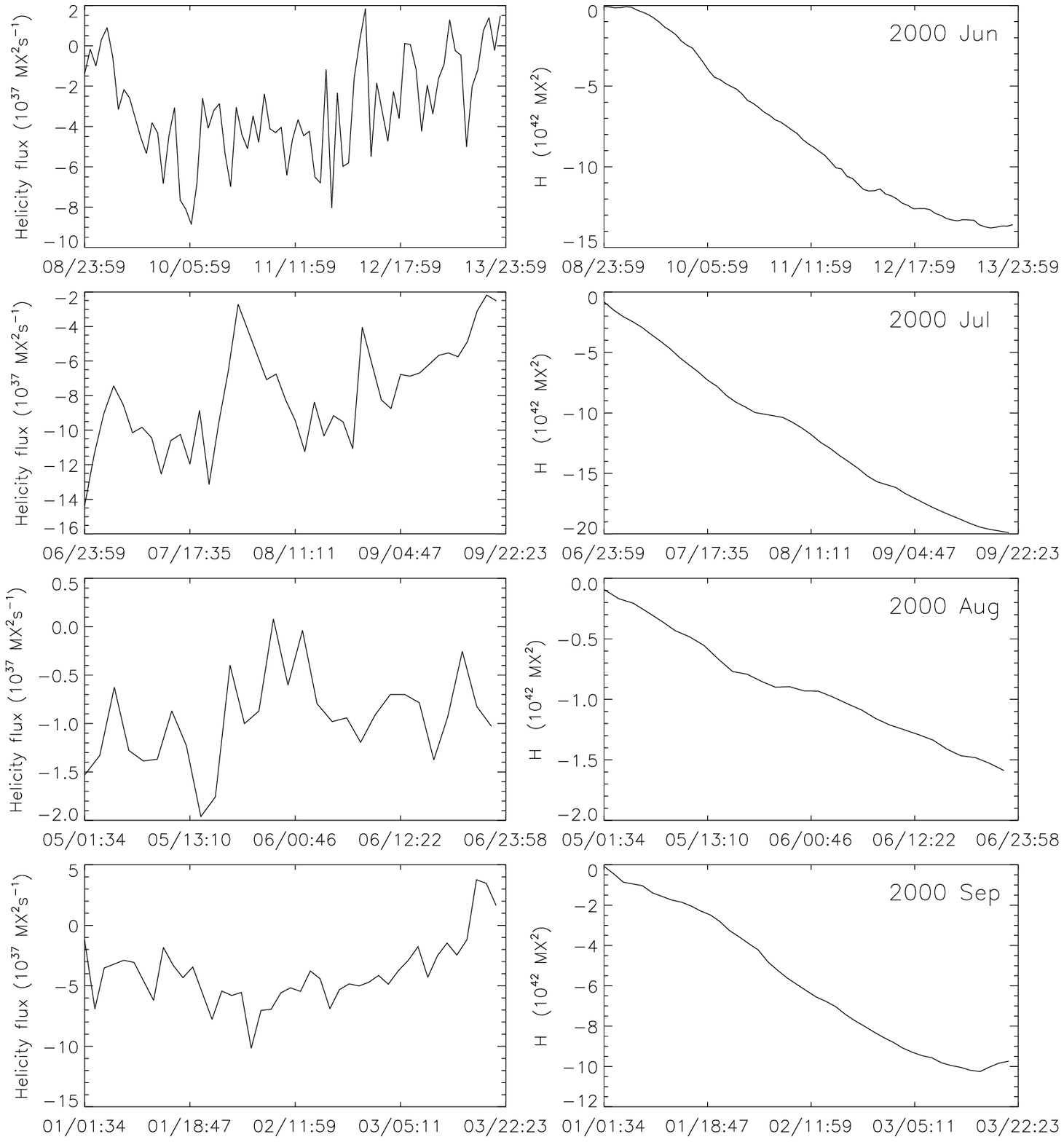}
\end{center}
\caption{The injection rate (left) and accumulation (right) of magnetic
helicity in the regions of Figure 1.}
\end{figure}

Figure 8 shows the rate of change and accumulation of magnetic
helicity inferred by MDI magnetograms relative to the large-scale
soft X-ray loop region of Figure 1. The accumulation of magnetic
helicity has been calculated using the local correlative tracking
(LCT) method (Chae, 2001) as the regions are located near the center
of the solar disk. Over all of the time intervals, it has been found
monotonic increase with time of negative helicity transfer from the
sub-atmosphere to the corona. The accumulated helicity is
-13.5$\times10^{42}Mx^2$ on 2000 June 9-13, -19.9$\times10^{42}Mx^2$
on 2000 July 7-19, -1.59$\times10^{42}Mx^2$ on 2000 August 5-6 and
-9.74$\times10^{42}Mx^2$ on 2000 September 1-3. The average
accumulation of magnetic helicity per day is
-3.60$\times10^{42}Mx^2$. As the contribution of September data is
disregarded due to the evanescence of the large-scale twisted soft
X-ray loops, the average accumulation of magnetic helicity is
-3.63$\times10^{42}Mx^2$ per day. It can be estimated that an order
of -4$\times10^{44}Mx^2$ magnetic helicity has been transferred into
the corona and contributed to the large-scale soft X-ray loops in
the period of 2000 June-September.

\begin{table*}[htb]
\begin{center}
\caption{The mean current helicity density and $\alpha$ parameter
inferred from vector
 magnetograms of active regions observed at Huairou relative to Figure 1.\label{tbl-2}}
\begin{tabular}{ccccc}
\tableline\tableline  date  & AR & lat (deg)&
$\overline{h_c}(10^{-2}G^2/m)$ & $\overline{\alpha}(10^{-7}/m)$\\
\tableline
2000 Jun 12-15 & 9033 & 15.8 & -0.329 &-0.209 \\
2000 Jul 06 & 9070 & 10.1 & -0.184 &-0.413 \\
2000 Aug 09-11 & 9114 & 10.3 & -0.085 &-0.125 \\
2000 Aug 31 - Sep 06 & 9149 & 6.7 & -0.059 &-0.113 \\
\tableline
\tableline
\end{tabular}
\end{center}
\end{table*}


Table 2 shows the the mean current helicity density and force free
$\alpha$ parameter of solar active regions calculated by vector
magnetograms observed at Huairou Solar Observing Station, National
Astronomical Observatories of China.  These active regions show the
negative sign of current helicity and they are same sign with
accumulated magnetic helicity. Active region NOAA 9033 was a fast
developing active region on 2000 June  13 in the right of Figure 1,
it was NOAA 9070 in the next solar rotation and it became the
large-scale enhanced magnetic networks in the magnetogram of August
6. Active region NOAA 9114 located in the left of magnetogram on
August 6 and active region NOAA 9149 near right bottom on September
3 in Figure 1. It can be estimated that the large-scale soft X-ray
loops are mainly contributed from active region NOAA 9033 and its
following rotated regions in the solar disk, while the contribution
from other active regions, such as NOAA9114 and 9149, also cannot be
neglected probably. This shows the consistence between the
accumulation of magnetic helicity and remained handedness of
magnetic field in the solar atmosphere.

\begin{table*}[htb]
\begin{center}
\caption{The mean current helicity density and $\alpha$ parameter
inferred from vector magnetograms of active regions observed at
Huairou and Mitaka relative to Figure 3.\label{tbl-3}}
\begin{tabular}{ccccccc}
\tableline\tableline  date  & AR &  time (UT) & lat (deg)& lon (deg)&
$\overline{h_c}(10^{-2}G^2/m)$ & $\overline{\alpha}(10^{-7}/m)$\\
\tableline
1992 Jun 24  & 7201 &02:48:30& 18.3 & 47.7 &1.591 & 0.683 \\
1992 Jul  & 7227& no data &   &   &   &   \\
1992 Aug 10  (J) & EH &01:23:31 & 24.4 &-36.8 & 0.366 & 0.228 \\
1992 Oct  & 7299 & no data  &   &  &  &   \\
\tableline
\tableline
\tablenotetext{}{$J$ is inferred by the vector magetogram on Aug-10,
1992 observed at National Astronomical Observatory of Japan in Mitaka and $EH$
indecats the Enhance network.\\}
\end{tabular}
\end{center}
\end{table*}


Table 3 shows the mean current helicity density and force free
$\alpha$ parameter of corresponding active regions in the target
region of Figure 3, which are inferred from vector magnetograms at
Huairou Solar Observing Station and National Astronomical
Observatory of Japan. It is found the positive sign of these
helicity parameters and they are consistent with the right
handedness of large-scale soft X-ray loops in Figure 3.

\begin{figure}
\begin{center}
\epsscale{0.90}
\plotone{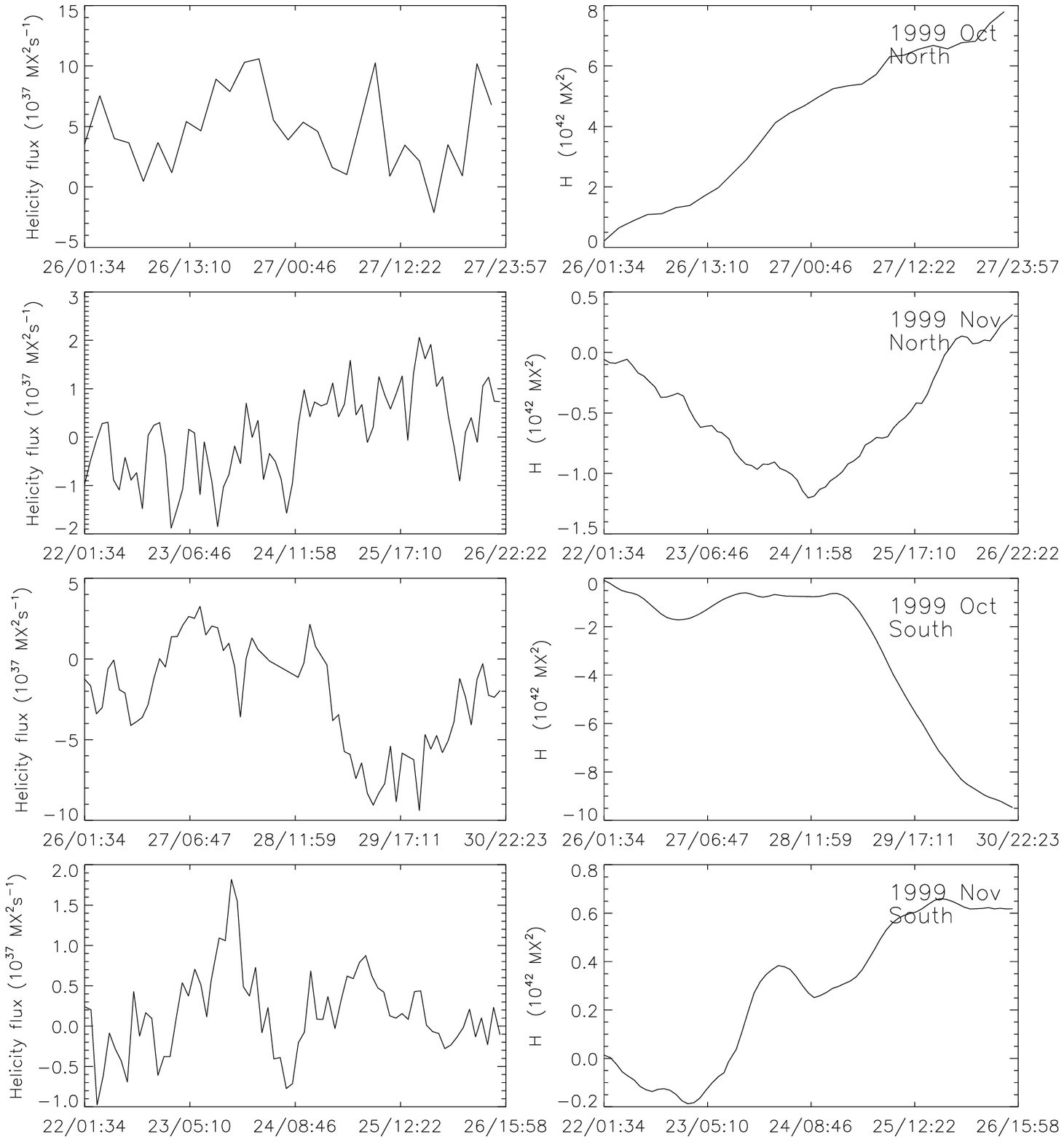}
\end{center}
\caption{The injection rate (left) and accumulation (right) of magnetic
helicity in the regions of both hemispheres marked by boxes  Figure 4.}
\end{figure}

\begin{table*}[htb]
\begin{center}
 \caption{The mean current helicity density and $\alpha$ parameter inferred from vector
 magnetograms of active regions observed at Huairou relative to Figure 5.\label{tbl-4}}
\begin{tabular}{cccccccc}
\tableline\tableline  hemisph& date  & AR &time (UT) & lat (deg)& lon (deg)&
$\overline{h_c}(10^{-2}G^2/m)$ & $\overline{\alpha}(10^{-7}/m)$\\
\tableline
 North & 2000 Nov  & no &   &   &  &  &  \\
 North & 2000 Dec 15  & 9272 & 03:39:29 & 20.9 & -18.9 &-0.0248 &-0.101 \\
 \tableline
 South & 2000 Nov 17  &  9231 &          & -29.3 & -33.3 &-0.138 &-0.128 \\
 South & 2000 Dec 15  &  9264 &          & -30.8 &-15.9 & 0.0797 & 0.0993 \\
\tableline
\tableline
\end{tabular}
\end{center}
\end{table*}

\begin{figure}
\begin{center}
\epsscale{0.90}
\plotone{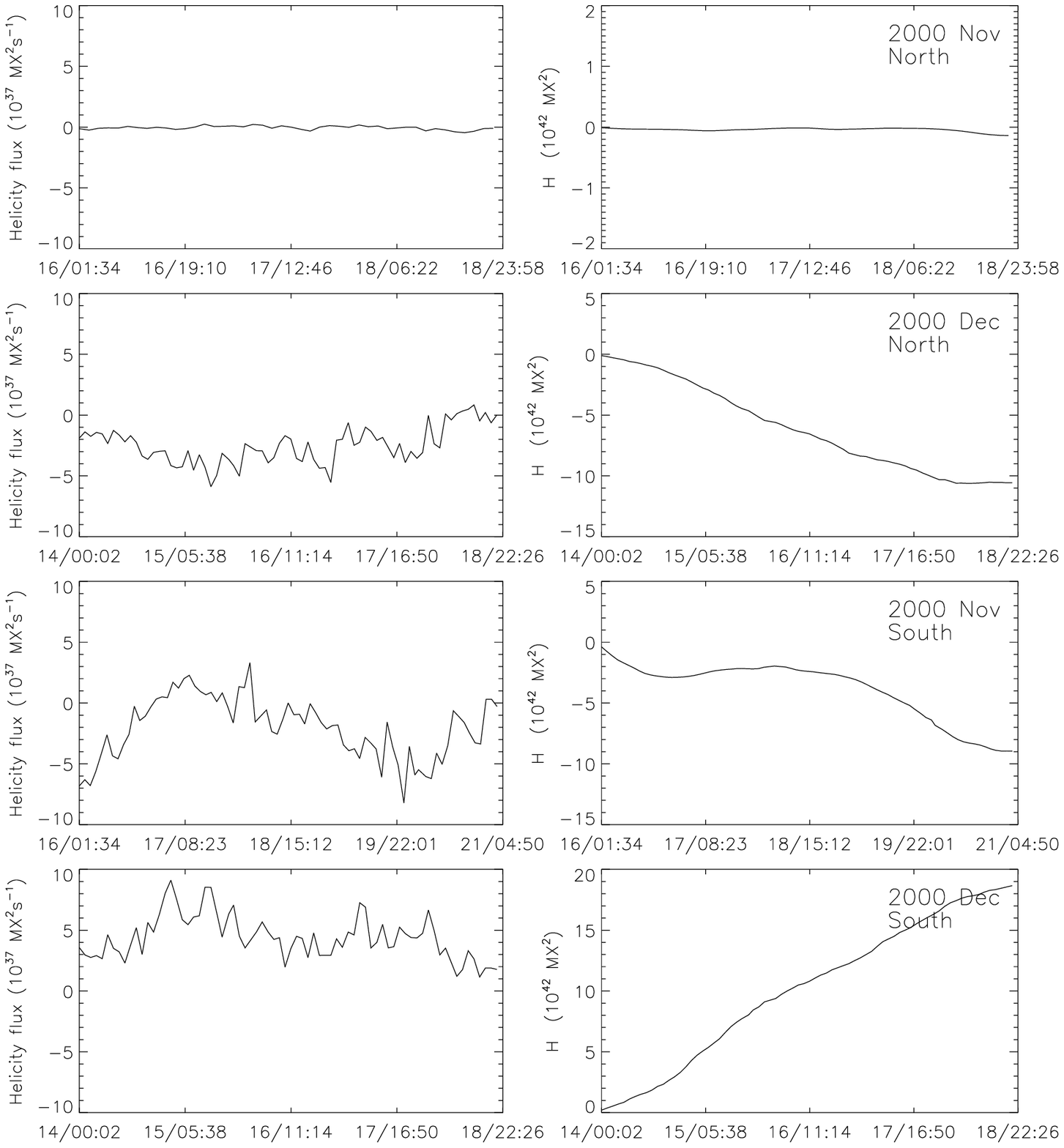}
\end{center}
\caption{The injection rate (left) and accumulation (right) of magnetic
helicity in the regions of both hemispheres marked by boxes  Figure 5.}
\end{figure}

A similar case on the same sign of magnetic helicity of solar active
regions in the northern and southern hemispheres can be found in
2003 October and November. These active regions produced a notable
amount of unexpected eruptive events (Zhang, et al., 2003; Berlicki,
et al., 2006; Guo et al., 2006; Li, et al., 2006; Mandrini, et al.,
2006; Uddin, et al., 2006; Zanna, et al., 2006). Liu and Zhang
(2006) found that about -6$\times10^{43}{Mx}^2$ magnetic helicity
has been transported from the subatmosphere into the corona in AR
10488 (L288,N8) in 2003 October 26 - 31. AR 10486 (L293, S15) has
been formed in the southern hemisphere. Zhang et al. (2008) detected
that -5$\times10^{43}Mx^2$ magnetic helicity from the active region
AR 10486 with the strong anticlockwise rotation of sunspots in 2003
Oct. 25 - 30. The mean current helicity of these active regions
shows the negative sign. This provides an interesting observational
evidence that one sign of magnetic helicity occurs in the whole of
the Sun. It means that the hemispheric sign rule can be disobeyed in
some periods of the solar cycle. It is also consistent with the
synthetic analysis of magnetic field in these active regions by Zhou
et al., (2007). This finding can be confronted with the account of
the total magnetic helicity injection over the whole solar cycle
presented by Georgoulis et al. (2009).

Figures 9 and 10 show the change rate and accumulation of magnetic
helicity inferred by MDI magnetograms relative to the large-scale
soft X-ray loop region in Figure 4 and 5, and also the corresponding
contribution in the solar disk in the previous solar rotation. The
accumulation of magnetic helicity is  calculated as the target
regions are located near the center of solar disk. It is found that
the accumulation process of magnetic helicity in these regions is
more complex than that of monotone variation in Figures 1 and 3.

Figure 9 shows the basic accumulation of positive magnetic helicity
in the previous solar rotation of the target region in the northern
hemisphere ($8\times 10^{42}Mx^2$ transports on 1999 October 26-27),
even if the sign change of accumulated magnetic helicity on November
22-24, while the increase of positive helicity in the target region
of southern hemisphere occurred on November 23-25.

Figure 10 shows about $-8.9\times 10^{42}Mx^2$ magnetic helicity
accumulated in the previous solar rotation of target region on 2000
November 16-21 in the southern hemisphere, {\bf even if about
$1.87\times 10^{43}Mx^2$ helicity had accumulated on 2000 December
16-18.  Figure 10 also shows that, in the northern hemisphere, the
accumulated helicity was about $-1.4\times 10^{41}Mx^2$ on 2000
November 16-18 and  $-1.1\times 10^{43}Mx^2$ on 2000 December
14-18.}

Table 4 displays the mean current helicity
density and and force free $\alpha$ parameter of corresponding
target active regions in Figure 5 and the region in the southern
hemisphere in the previous solar rotation, which are inferred from
vector magnetograms at Huairou Solar Observing Station.  It is found
that the mean current helicity parameters show negative sign
inferred from vector magnetograms  in active region NOAA 9231 on
2000 November 17 and positive sign in the corresponding region NOAA
9264 in the next solar  rotation on 2000 December 15. As comparing
with Figure 10, it is found the sign difference of mean current
helicity parameters is consistent with the accumulation of magnetic
helicity in the target regions in Figure 5. This means that the
formation of reverse handedness of large-scale soft X-ray loops is
contributed by the transfer of magnetic helicity from the
subatmosphere and is a relative complex process.

\section{Discussion}

In this paper, we have demonstrated the large scale soft X-ray loops
visible in soft X-ray images and their relationship with the
magnetic helicity. Thereby, we have presented examples of that the
magnetic chirality holds the same handedness with transfer upwards
from the subatmosphere. It is consistent with the morphologically
same handedness of soft X-ray loops in both hemispheres as shown in
Figures 2-5. Furthermore, we have demonstrated examples of the
statistical imbalance in the opposite handedness of soft X-ray loops
and the magnetic helicity in some phase of solar cycles (such as, in
1992 June-October and 2003 October - November). Thus, we have
observed the cases of the same sign of the magnetic helicity across
the equator simultaneously, contrary to the usual occurrence of the
opposite sign.

The large-scale soft X-ray loops are connected with solar active
regions and enhanced magnetic networks, and we have shown the
statistical relation of their handedness with the the hemispheric
rule for the helicity in the both hemispheres. In some years, such
as in 1997 and 1998 in our statistical analysis, this correlation
appears very high. The fraction of the soft X-ray loops following
the hemispheric rule is slightly different from that of the current
helicity inferred from vector magnetograms (e.g., Pevtsov et al.,
1995; Bao and Zhang, 1998; Hagino and Sakurai, 2005; Zhang et al.,
2010), due to the difference in the nature of this parameter, while
it may also provide some message on the mechanism of generation of
magnetic field in the subatmosphere and the reversal of helicity of
magnetic field in the solar atmosphere at some latitudes and periods
of the solar cycle.

The statistical distribution of the mean current helicity density of
solar active regions presented in Figure 2 of Zhang et al. (2010)
shows evidence on the imbalance of helicity at the decaying phase of
solar cycles 22 and 23. One can see domains in latitude and time of
the reversed with respect to the statistical hemispheric rule sign
at the beginning as well as end of the butterfly wings. It is
consistent with the results in this paper, although we focus on
morphological and statistical analysis of the handedness of
large-scale soft X-ray loops of active regions and enhanced
networks. We propose this phenomenon as the penetration of the
activity wave from one hemisphere into the other one, a sort of
trespassing into the $"wrong"$ hemisphere with respect to the
average sign of the helicity.

The magnetic helicity can be considered as a measure of mirror
asymmetry of solar magnetic fields. They are generated, according to
the mean-field solar dynamo model, due to the effects of solar
differential rotation and the action of Coriolis force on turbulent
motions of plasma in the solar convection zone (e.g. Berger and
Ruzmaikin, 2000; Kuzanyan et al., 2000; Kleeorin et al., 2003;
Choudhuri 2003; Choudhuri et al., 2004; Zhang et al., 2006).
However, the issue of occurrence of the same sign components of
magnetic helicity in the both hemisphere have not been studies
sufficiently.

The Sun is an open system for the magnetic helicity. The eruption of
flare-CMEs brings the magnetic helicity from solar atmosphere into
the interplanetary space (Zhang and Low, 2005), which probably
causes the imbalance of remaining magnetic helicity in the
convection zone in the both hemispheres. Due to the eruption
process, the balance of helicity in both the hemispheres can be
seriously distorted, even if the dynamo and turbulent convection
mechanism produce helicity of opposite sign across the equator and
absolute amounts of helicity are comparable. This disbalance of
helicity can be compensated due to the large-scale helicity
transport from one hemisphere to another one.
The possible agent of such transfer could be the trans-equatorial
magnetic field inside the solar convection zone.

Within the framework of the solar dynamo model, we can interpret the
trans-equatorial interaction of dynamo waves as a proxy of global
modulation of the solar activity such as Gleissberg cycle. However,
simple 1D models of magnetic field interaction across the equator
(such as e.g, Galitskii et al., 2005) can hardly interact
effectively enough for quantitatively realistic times, which for
Gleissberg cycle is of order 100 yr. Hereby we may suggest an
additional mechanism of interaction of dynamo waves across the
equator by means of transfer of twist and helicity which may occur
more effective than one of magnetic field itself. We also may
suggest for future studies to investigate such possible mechanism of
interaction at atmospheric and subatmospheric level in the Sun,
probably by means of Alfvenic waves.

Further development of the mechanism on the generation of local
magnetic helicity in the convection zone, as well as the global
evolution of helicity with the solar cycle and its parts which are
symmetric and anti-symmetric with respect to the equator, still
remain a challenge for theory.

For the forthcoming studies of basic properties of the solar dynamo,
the symmetric and anti-symmetric parts of magnetic helicity of solar
active regions in the both hemispheres and their evolution with the
solar cycle still need to be observationally confirmed in detail.

\medskip
\medskip


{\bf Acknowledgments}
The authors would like to Prof. Arnab Rai
Choudhuri for his comments and suggestions on the manuscript. This
study is supported by the grants -- National Basic Research Program
of China: 2006CB806301, National Natural Science Foundation of
China: 10611120338, 10473016, 10673016, 60673158, 10733020 and
Important Directional Project of Chinese Academy of Sciences:
KLCX2-YW-T04. This work has been carried out within the framework of
the joint Chinese-Russian collaborative project. We would like to
acknowledge support from NNSF of China and RFBR of Russia under
grant 08-02-92211, and also RFBR grant 10-02-00960-a. KK and DS
would like to thank Visiting Professorship program of Chinese
Academy of Sciences and NAOC for support of their visits to Beijing.

\end{document}